\documentclass[%
 reprint,
 amsmath,amssymb,
 aps,
]{revtex4-2}

\usepackage{graphicx}
\usepackage{dcolumn}
\usepackage{bm}
\usepackage{xcolor}
\usepackage{algorithm} 
\usepackage{algpseudocode} 
\usepackage{tabularx}

\begin{document}


\title{Copula-based analytical results of horizontal visibility graphs for correlated time series}

\author{Jeong-Min Lee}
\affiliation{Department of Physics, The Catholic University of Korea, Bucheon, Republic of Korea}

\author{Hang-Hyun Jo}
\email{h2jo@catholic.ac.kr}
\affiliation{Department of Physics, The Catholic University of Korea, Bucheon, Republic of Korea}

\date{\today}

\begin{abstract}
The visibility graph (VG) algorithm and its variants have been extensively studied in the time series analysis as they transform the time series into the network of nodes and links, enabling to characterize the time series in terms of network measures such as degree distributions. Despite numerous practical applications of VGs in various disciplines, analytical, rigorous understanding of VGs for the correlated time series is still far from complete due to the lack of mathematical tools for modeling the correlation structure in the time series in a tractable form. In this work, we adopt the Farlie-Gumbel-Morgenstern (FGM) copula method to derive the analytical solutions of degree distributions of the horizontal visibility graph (HVG) and its directed version (DHVG) for the correlated time series. Our analytical results show exactly how the correlation between consecutive data points affects the degree distributions of HVGs and DHVGs up to the first order of the correlation parameter in the FGM copula. Thus, our findings shed light on the rigorous understanding of the VG algorithms.
\end{abstract}

\maketitle

\section{Introduction}\label{sec:intro}

Nonlinear time series analysis has been extensively studied for several decades due to its importance in understanding the dynamics of complex systems~\cite{Kantz2004Nonlinear}. In recent years, the emerging field of network science~\cite{Albert2002Statistical, Barabasi2016Network, Newman2018Networks, Menczer2020First} has been combined with the time series analysis to broaden the scope of both fields, e.g., see Ref.~\cite{Zou2019Complex} and references therein. Among a number of network approaches to time series analysis, we focus on the visibility graph (VG) algorithm that transforms the time series into networks of nodes and links~\cite{Lacasa2008Time, Nunez2012Visibility, Luque2016Entropy, Azizi2024Review}. That is, each data point in the time series is denoted by a node; the $i$th datum $x_i$ is observed at the time point $t_i$. The link between two nodes is created whenever they ``see" each other. Precisely, a node $i$ and another node $j$ are connected when the following condition is satisfied:
\begin{align}
    \frac{x_i-x_k}{t_i-t_k}<\frac{x_i-x_j}{t_i-t_j}
    \label{eq:VG_define}
\end{align}
for all time points $t_k$ with $t_i<t_k<t_j$. One of variants of VGs is the horizontal visibility graph (HVG)~\cite{Luque2009Horizontal, Lacasa2010Description, Lacasa2014Degree} in which the link between two nodes $i$ and $j$ is created whenever the following condition is satisfied:
\begin{align}
    x_k<\min\{x_i,x_j\}
    \label{eq:HVG_define}
\end{align}
for all $t_k$ with $t_i<t_k<t_j$. Its directed version, namely, the DHVG, has also been studied~\cite{Lacasa2012Time, Lacasa2014Degree}; the directed link in the DHVG is created from $i$ to $j$ by Eq.~\eqref{eq:HVG_define} but only when $t_i<t_j$. Then various measures developed in network science, such as degree distributions, can be employed to characterize the time series.

Despite numerous practical applications of VGs in diverse fields~\cite{Nunez2012Visibility, Luque2016Entropy, Zou2019Complex}, analytical, rigorous understanding of the VG algorithms is still lacking except for a few results. For the simplest, uncorrelated time series, the degree distribution of the HVG has been analytically derived as~\cite{Luque2009Horizontal}
\begin{align}
    P^{\rm und}(k)=\frac{1}{3}\left(\frac{2}{3}\right)^{k-2}\ \textrm{for}\ k=2,3,\ldots,
    \label{eq:Pk_und_uncorrel}
\end{align}
where the superscript ``und'' means the original, undirected HVG compared to the DHVG. The indegree and outdegree distributions of the DHVG for the uncorrelated time series have been obtained as~\cite{Lacasa2012Time}
\begin{align}
    P^{\rm in}(k)= P^{\rm out}(k) =\left(\frac{1}{2}\right)^{k}\ \textrm{for}\ k=1,2,\ldots.
    \label{eq:Pk_out_uncorrel}
\end{align}
There also exist some exact results of another kind of the HVG, namely, the limited penetrable HVG, derived for uncorrelated time series~\cite{Wang2018Exact}. In contrast to the uncorrelated time series, most time series observed in complex systems are correlated, requiring us to analyze the VGs for the correlated time series. A theoretical approach to the HVG and DHVG for correlated time series has been studied in Refs.~\cite{Lacasa2010Description, Lacasa2014Degree}; e.g., the concrete solutions of degree distributions were obtained for the Ornstein–Uhlenbeck (OU) process~\cite{VanKampen2007Stochastic} up to $k=3$. The OU process is Markovian in the sense that each datum in the time series is dependent only on its previous datum. However, the effects of such dependency on degree distributions of HVGs and DHVGs have not been fully understood, possibly due to the lack of mathematical tools for descrbing the dependency structure between data points in the time series.

In this work, we take an alternative approach to the analysis of HVGs and DHVGs for the correlated time series by adopting the copula method~\cite{Nelsen2006Introduction}. The copula method enables us to explicitly write down the joint probability distribution of two or more consecutive $x$ values in a tractable form. Such a joint probability distribution carries information on the correlation between those values. The copula method has been used in various disciplines such as finance~\cite{Embrechts2009Copulas}, time series analysis~\cite{Jo2019Analytically, Jo2019Copulabased, Yu2025Analysis}, astronomy~\cite{Takeuchi2010Constructing, Takeuchi2020Constructing}, biology~\cite{Ray2020CODC}, engineering~\cite{Horvath2020CopulaBased}, and network science~\cite{Jo2021Analytical, Jo2022Copulabased, Jo2023Copulabased}. Here we adopt the Farlie-Gumbel-Morgenstern (FGM) copula among many others~\cite{Schucany1978Correlation, Huang1984Correlation, Nelsen2006Introduction}, which contains a single correlation parameter $\rho$, controlling the degree of correlation between two consecutive $x$ values. We first propose a model for the correlated time series using the FGM copula, and then derive the analytical solutions of degree distributions of the HVG and DHVG for the model time series as a function of $\rho$ up to $k=4$ for the HVG and $k=3$ for the DHVG, irrespective of the functional form of the distribution of $x$ values. We analytically and numerically find that the role of $\rho$ in the degree distribution varies with $k$, namely, it can be either a constant, increasing, decreasing, or non-monotonic function of $\rho$ depending on $k$.

Our paper is organized as follows: In Sec.~\ref{sec:model} we introduce a model generating the correlated time series using the FGM copula, following the method in Ref.~\cite{Jo2019Copulabased}. In Sec.~\ref{sec:DHVG} we derive the analytical results of the indegree and outdegree distributions of the DHVG for the model time series up to $k=4$ and up to the first order of $\rho$. These analytical results are found to be in good agreement with numerical simulations. We then analyze the HVG for the same model time series in Sec.~\ref{sec:HVG} to find that the analytical results are supported by the numerical simulations. Finally, we conclude our paper in Sec.~\ref{sec:concl}.

\section{Correlated time series model}\label{sec:model}

We propose a model generating the correlated time series in $n$ discrete times $\{x_i\}_{i=1,\ldots,n}$, where the subscript $i$ denotes the time point, i.e., $t_i=i$. To implement the correlation between two consecutive values, say $x_i$ and $x_{i+1}$, we adopt the Farlie-Gumbel-Morgenstern (FGM) copula~\cite{Nelsen2006Introduction}. Precisely, when the distribution of $x$ values is denoted by $P(x)$, the joint probability distribution of $x_i$ and $x_{i+1}$ is given by
\begin{align}
	P(x_i, x_{i+1})=P(x_i)P(x_{i+1}) \left[1+\rho f(x_i)f(x_{i+1})\right],
	\label{eq:Pxx}
\end{align}
where 
\begin{align}
    f(x) \equiv 2F(x)-1,\ F(x) \equiv \int_{-\infty}^{x} {\rm d}x' P(x').
\end{align}
Here $F(x)$ is the cumulative distribution function of $P(x)$. The correlation parameter $\rho\in[-1,1]$ controls the correlation between $x_i$ and $x_{i+1}$. Using Eq.~\eqref{eq:Pxx} one can define the memory coefficient $M$~\cite{Jo2019Analytically, Goh2008Burstiness} as follows:
\begin{align}
    M\equiv \frac{\langle x_ix_{i+1}\rangle-\langle x\rangle^2}{\sigma^2},
    \label{eq:M_define}
\end{align}
where
\begin{align}
    \langle x_ix_{i+1}\rangle\equiv \int_{-\infty}^{\infty} {\rm d}x_i\int_{-\infty}^{\infty} {\rm d}x_{i+1}x_ix_{i+1}P(x_i, x_{i+1}),
\end{align}
and $\sigma^2$ denotes the variance of $x$ values. $M$ has a value in the range of $[-1,1]$; the positive $M$ means that a large (small) value tends to follow a large (small) value. The negative $M$ implies the opposite tendency, while $M\approx 0$ indicates no correlations between two consecutive values. Plugging Eq.~\eqref{eq:Pxx} into Eq.~\eqref{eq:M_define}, it turns out that $M$ in Eq.~\eqref{eq:M_define} is a linear function of $\rho$:
\begin{align}
    M=a\rho,\ a\equiv \frac{1}{\sigma^2}\left[\int_{-\infty}^\infty {\rm d}x x  P(x)f(x)\right]^2.
    \label{eq:M_arho}
\end{align}
Since $|\rho|\leq 1$, we get $|M|\leq a$. The upper bound of $a$ for any $P(x)$ has been proven to be $1/3$~\cite{Schucany1978Correlation}. We also remark that the FGM copula in Eq.~\eqref{eq:Pxx} is symmetric with respect to the exchange of two variables, namely,
\begin{align}
    P(x_i, x_{i+1})=P(x_{i+1}, x_i),
    \label{eq:time_reverse}
\end{align}
which implies the time-reversal symmetry of the FGM copula.

Let us now define the correlated time series model. We first randomly draw $x_1$ from the given $P(x)$ and the next value $x_2$ is randomly drawn from the conditional probability distribution $P(x_2|x_1)$. Then $x_3$ is drawn from $P(x_3|x_2)$, and so on. In general, $P(x_{i+1}|x_i)$ is given from Eq.~\eqref{eq:Pxx} as
\begin{align}
	P(x_{i+1}|x_i)=P(x_{i+1}) \left[1+\rho f(x_i)f(x_{i+1})\right].
	\label{eq:Px|x}
\end{align}
This procedure is repeated until $n$ values are drawn to generate the time series $\{x_1,\ldots,x_n\}$.

\begin{figure*}[!t]
\centering
\includegraphics[width=0.8\textwidth]{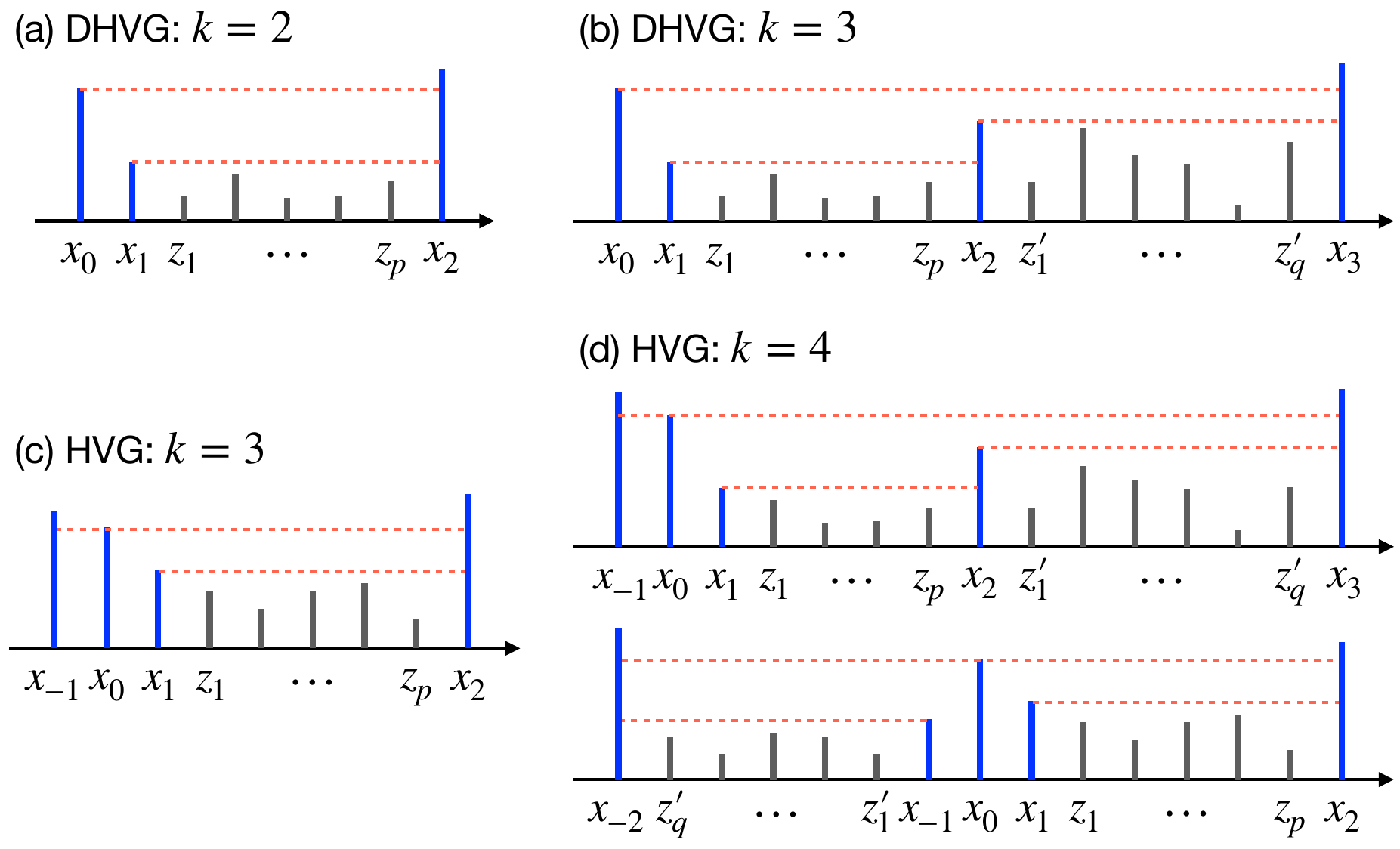}
\caption{Schematic diagram for calculating the outdegree distribution $P^{\rm out}_{\rho}(2)$ (a) and $P^{\rm out}_{\rho}(3)$ (b) for the directed horizontal visibility graph (DHVG) and the degree distribution $P^{\rm und}_{\rho}(3)$ (c) and $P^{\rm und}_{\rho}(4)$ (d) for the horizontal visibility graph (HVG). In all panels, red horizontal dashed lines are for guiding eyes.
}
\label{fig:scheme}
\end{figure*}

\section{Directed horizontal visibility graph}\label{sec:DHVG}

\subsection{Analysis}\label{subsec:analysis}

We derive the analytical solution of the indegree and outdegree distributions of the directed horizontal visibility graphs for the model time series presented in Sec.~\ref{sec:model}. Thanks to the time-reversal symmetry of the FGM copula in Eq.~\eqref{eq:time_reverse}, indegree and outdegree distributions are identical. Hence we focus on the derivation of the outdegree distribution hereafter.

We first study the probability of nodes having the outdegree of one, i.e., $k=1$, which happens whenever $x_i\leq x_{i+1}$, irrespective of other parts of the time series. As our model time series can be considered as a stationary stochastic process, we respectively denote $x_i$ and $x_{i+1}$ by $x_0$ and $x_1$ without loss of generality; in this subsection the subscripts $0$ and $1$ imply only the order of two $x$ values, not the real time. Then the case with $k=1$ occurs when $x_0\leq x_1$:
\begin{align}
    P^{\rm out}_{\rho}(1)=\int_{-\infty}^{\infty} {\rm d}x_0 P(x_0) \int_{x_0}^{\infty} {\rm d}x_1 P(x_1|x_0).
    \label{eq:P1_out_define}
\end{align}
Using $P(x_{i+1}|x_i)$ in Eq.~\eqref{eq:Px|x}, we obtain the exact solution as
\begin{align}
    P^{\rm out}_{\rho}(1)=\frac{1}{2}.
    \label{eq:P1_out_sol}
\end{align}
We remark that this solution has been derived for the entire range of $\rho$ and for the arbitrary functional form of $P(x)$. In addition, this result is the same as $P^{\rm out}(1)=1/2$ for the uncorrelated time series in Eq.~\eqref{eq:Pk_out_uncorrel}~\cite{Lacasa2014Degree}. From now on, we denote the outdegree distribution for the uncorrelated case by $P^{\rm out}_0(k)$ to be consistent with our definition of $P^{\rm out}_{\rho}(k)$.

Next, we consider the case with $k=2$, for which three ordered values $x_0,x_1,x_2$ satisfy the condition that $x_1<x_0\leq x_2$, as depicted in Fig.~\ref{fig:scheme}(a). Here there can be an arbitrary number of intermediate values, say $z_1,\ldots,z_p$, between $x_1$ and $x_2$ such that $z_i\leq x_1$ for all $i=1,\ldots,p$. These $z$ values are also called hidden variables in Ref.~\cite{Lacasa2014Degree}. One can write
\begin{align}
    P^{\rm out}_{\rho}(2)=\int_{-\infty}^{\infty} {\rm d}x_0 P(x_0) \int_{-\infty}^{x_0} {\rm d}x_1 P(x_1|x_0)\sum_{p=0}^\infty h_p(x_0,x_1),
    \label{eq:Pto2_define}
\end{align}
where for $p=0$
\begin{align}
    h_0(x_0,x_1)\equiv \int_{x_0}^{\infty} {\rm d}x_2 P(x_2|x_1),
    \label{eq:G0_define}
\end{align}
and for $p\geq 1$
\begin{align}
    h_p(x_0,x_1) &\equiv 
    \int_{-\infty}^{x_1} {\rm d}z_1 P(z_1|x_1)
    \left[\prod_{i=2}^p \int_{-\infty}^{x_1} {\rm d}z_i P(z_i|z_{i-1})\right] \notag\\
    &\times \int_{x_0}^{\infty} {\rm d}x_2 P(x_2|z_p).
    \label{eq:Gp_define}
\end{align}
Note that when $p=1$, the integrals in the brackets of Eq.~\eqref{eq:Gp_define} are ignored. Using Eq.~\eqref{eq:Px|x} and denoting $y_i\equiv F(x_i)$ and $\bar y_i\equiv 1-y_i$, we obtain for $p=0$
\begin{align}
    h_0(x_0,x_1)=\bar y_0[1+\rho y_0(2y_1-1)],
\end{align}
and for $p\geq 1$
\begin{align}
    h_p(x_0,x_1)=\bar y_0 y_1^p [1-\rho \bar y_1(y_0+y_1-p\bar y_1)]+\mathcal{O}(\rho^2),
\end{align}
which can be proven by induction. Here we have assumed that $|\rho|\ll 1$ to ignore higher order terms of $\rho$, otherwise $h_p(x_0,x_1)$ could have been of the order of $\rho^{p+1}$, though the higher-order terms of $\rho$ can be calculated in a similar manner. The summation of $h_p(x_0,x_1)$ reads
\begin{align}
    \sum_{p=0}^\infty h_p(x_0,x_1)=\frac{\bar y_0}{\bar y_1}-\rho \bar y_0 \bar y_1 (y_0-y_1)+\mathcal{O}(\rho^2).
    \label{eq:Gp_sum}
\end{align}
Plugging Eq.~\eqref{eq:Gp_sum} into Eq.~\eqref{eq:Pto2_define}, we finally obtain the analytical solution of $P^{\rm out}_{\rho}(2)$ for any $P(x)$ up to the first order of $\rho$:
\begin{align}
    P^{\rm out}_{\rho}(2) = \frac{1}{4}-\frac{1}{180}\rho +\mathcal{O}(\rho^2).
    \label{eq:P2_out_sol}
\end{align}
For the uncorrelated case with $\rho=0$, the result in Eq.~\eqref{eq:P2_out_sol} reduces to $P^{\rm out}_0(2)=1/4$ for the uncorrelated case in Eq.~\eqref{eq:Pk_out_uncorrel}~\cite{Lacasa2014Degree}. For $|\rho|\ll 1$, we find that $P^{\rm out}_{\rho}(2)$ is a decreasing function of $\rho$. 

\begin{widetext}
The case with $k=3$ is depicted in Fig.~\ref{fig:scheme}(b), where four ordered values $x_0,x_1,x_2,x_3$ satisfy the condition that $x_1<x_0\leq x_3$ and $x_1<x_2<x_0$, while there can be an arbitrary number of intermediate values, say $z_1,\ldots,z_p$, between $x_1$ and $x_2$ such that $z_i\leq x_1$ for all $i=1,\ldots,p$, as well as an arbitrary number of intermediate values, say $z'_1,\ldots,z'_q$, between $x_2$ and $x_3$ such that $z'_j\leq x_2$ for all $j=1,\ldots,q$. One can write
\begin{align}
    P^{\rm out}_{\rho}(3)=\int_{-\infty}^{\infty} {\rm d}x_0 P(x_0) \int_{-\infty}^{x_0} {\rm d}x_1 P(x_1|x_0)\sum_{p,q=0}^{\infty} h_{pq}(x_0,x_1),
    \label{eq:Pto3_define}
\end{align}
where we have for $p=q=0$
\begin{align}
    h_{00}(x_0,x_1)\equiv 
    \int_{x_1}^{x_0} {\rm d}x_2 P(x_2|x_1)
    \int_{x_0}^{\infty} {\rm d}x_3 P(x_3|x_2),
    \label{eq:G00_define}
\end{align}
for $p\geq 1$ and $q=0$
\begin{align}
    h_{p0}(x_0,x_1) &\equiv 
    \int_{-\infty}^{x_1} {\rm d}z_1 P(z_1|x_1)
    \left[\prod_{i=2}^p \int_{-\infty}^{x_1} {\rm d}z_i P(z_i|z_{i-1})\right] \int_{x_1}^{x_0} {\rm d}x_2 P(x_2|z_p)  \int_{x_0}^{\infty} {\rm d}x_3 P(x_3|x_2),
    \label{eq:Gp0_define}
\end{align}
for $p=0$ and $q\geq 1$
\begin{align}
    h_{0q}(x_0,x_1) &\equiv 
    \int_{x_1}^{x_0} {\rm d}x_2 P(x_2|x_1) \int_{-\infty}^{x_2} {\rm d}z'_1 P(z'_1|x_2)
    \left[\prod_{j=2}^q \int_{-\infty}^{x_2} {\rm d}z'_j P(z'_j|z'_{j-1})\right] \int_{x_0}^{\infty} {\rm d}x_3 P(x_3|z'_q),
    \label{eq:G0q_define}
\end{align}
and for $p,q\geq 1$
\begin{align}
    h_{pq}(x_0,x_1) &\equiv 
    \int_{-\infty}^{x_1} {\rm d}z_1 P(z_1|x_1)
    \left[\prod_{i=2}^p \int_{-\infty}^{x_1} {\rm d}z_i P(z_i|z_{i-1})\right] \int_{x_1}^{x_0} {\rm d}x_2 P(x_2|z_p) \notag\\
    &\times \int_{-\infty}^{x_2} {\rm d}z'_1 P(z'_1|x_2)
    \left[\prod_{j=2}^q \int_{-\infty}^{x_2} {\rm d}z'_j P(z'_j|z'_{j-1})\right] \int_{x_0}^{\infty} {\rm d}x_3 P(x_3|z'_q).
    \label{eq:Gpq_define}
\end{align}
\end{widetext}
Note that when $p=1$ or $q=1$, the integrals in the brackets of Eqs.~\eqref{eq:Gp0_define}--\eqref{eq:Gpq_define} are ignored. Using Eq.~\eqref{eq:Px|x}, we get for $p=q=0$
\begin{align}
    h_{00}(x_0,x_1)=\bar y_0[I_1+\rho(\bar y_0-2y_1)(I_1-2I_2)]+\mathcal{O}(\rho^2),
    \label{eq:G00_sol}
\end{align}
for $p\geq 1$ and $q=0$
\begin{align}
    &h_{p0}(x_0,x_1)=\bar y_0y_1^p\{I_1-\rho[(y_0-\bar y_1^2)I_1 + 2(\bar y_0-y_1)I_2 \notag\\
    & -p\bar y_1^2 I_1]\}+\mathcal{O}(\rho^2),
    \label{eq:Gp0_sol}
\end{align}
for $p=0$ and $q\geq 1$
\begin{align}
    &h_{0q}(x_0,x_1)=\bar y_0 \{I_{q+1}+\rho[(\bar y_0-2y_1+q)I_{q+1}\notag\\ 
    & -(\bar y_0-4y_1+2q+2)I_{q+2} + (q+1)I_{q+3}]\}+\mathcal{O}(\rho^2),
    \label{eq:G0q_sol}
\end{align}
and for $p,q\geq 1$
\begin{align}
    &h_{pq}(x_0,x_1)=\bar y_0y_1^p \{I_{q+1}-\rho[(y_0-\bar y_1^2-q)I_{q+1}\notag\\
    &+(\bar y_0+2\bar y_1+2q)I_{q+2} - (q+1)I_{q+3} - p\bar y_1^2I_{q+1}]\}+\mathcal{O}(\rho^2).
    \label{eq:Gpq_sol}
\end{align}
Here 
\begin{align}
    I_m\equiv \frac{y_0^{m}-y_1^{m}}{m}\ \textrm{for}\ m=1,2,\ldots,
\end{align}
and $|\rho|\ll 1$ has been assumed to ignore the higher-order terms of $\rho$. The summation of $h_{pq}(x_0,x_1)$ reads
\begin{align}
    &\sum_{p,q=0}^{\infty} h_{pq}(x_0,x_1) = \frac{\bar y_0}{\bar y_1}\log\left(\frac{\bar y_1}{\bar y_0}\right)\notag\\ 
    &+\rho \frac{\bar y_0}{6\bar y_1}\left[-\bar y_0^3-9\bar y_0\bar y_1^2+10\bar y_1^3+6\bar y_1^3\log \left(\frac{\bar y_0}{\bar y_1}\right)\right] +\mathcal{O}(\rho^2).
    \label{eq:Gpq_sum}
\end{align}
Plugging Eq.~\eqref{eq:Gpq_sum} into Eq.~\eqref{eq:Pto3_define}, we obtain the analytical solution of $P^{\rm out}_{\rho}(3)$ for any $P(x)$ up to the first order of $\rho$ as 
\begin{align}
    P^{\rm out}_{\rho}(3)=\frac{1}{8}+\frac{29}{5400}\rho +\mathcal{O}(\rho^2).
    \label{eq:P3_out_sol}
\end{align}
For the uncorrelated case with $\rho=0$, the result in Eq.~\eqref{eq:P3_out_sol} reduces to $P^{\rm out}_0(3)=1/8$ for the uncorrelated case in Eq.~\eqref{eq:Pk_out_uncorrel}~\cite{Lacasa2014Degree}. For $|\rho|\ll 1$, we find that $P^{\rm out}_{\rho}(3)$ is an increasing function of $\rho$. 

\begin{figure*}[!t]
\centering
\includegraphics[width=0.85\textwidth]{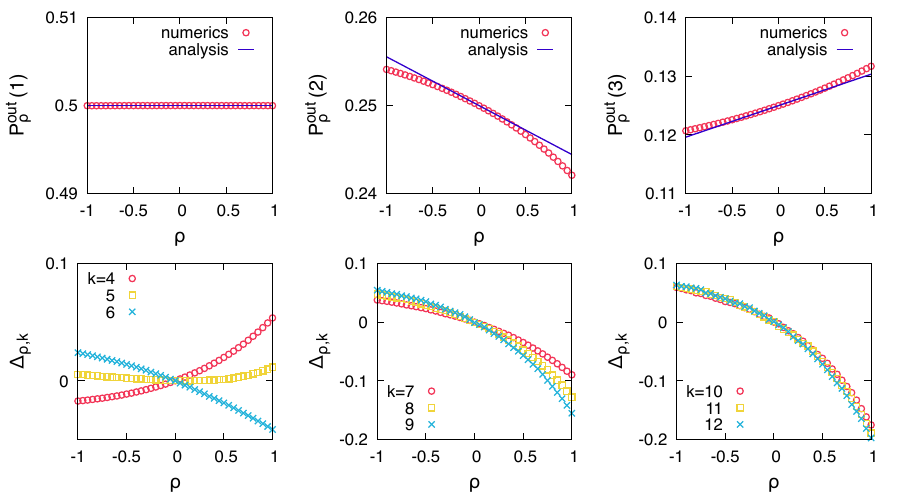}
\caption{Numerical results (symbols) of the outdegree distribution, i.e., $\tilde P^{\rm out}_{\rho}(k)$ for $k=1,\ldots,12$, of the directed horizontal visibility graphs derived from the model time series. For the simulation, we generate $100$ different time series with $n=10^7$ using $P(x)=1$ defined over $0\leq x<1$ for various values of $\rho$. In upper panels solid lines respectively show the analytical solutions of $P^{\rm out}_{\rho}(1)$ in Eq.~\eqref{eq:P1_out_sol}, $P^{\rm out}_{\rho}(2)$ in Eq.~\eqref{eq:P2_out_sol}, and $P^{\rm out}_{\rho}(3)$ in Eq.~\eqref{eq:P3_out_sol} as functions of $\rho$. In lower panels $\Delta_{\rho,k}$ in Eq.~\eqref{eq:Delta} is plotted for $k=4,\dots,12$.
}
\label{fig:DHVG}
\end{figure*}

Finally, we remark that the analytical solutions of $P^{\rm out}_{\rho}(k)$ for $k\geq 4$ can be obtained in a similar way.

\subsection{Numerical simulation}\label{subsec:DHVG_numerics}

To generate the correlated time series, we adopt the copula-based method proposed in Ref.~\cite{Jo2019Copulabased}. We first draw a number $x_1$ randomly from the given $P(x)$. The next number $x_2$ is drawn from the conditional distribution $P(x_2|x_1)$ given in Eq.~\eqref{eq:Px|x}, and so on. Precisely, for a given $x_i$ and a random number $r$ drawn from a uniform distribution defined over $[0,1)$, the value $x_{i+1}$ is given as 
\begin{align}
    x_{i+1}=F^{-1}\left[\frac{u_i-1+\sqrt{(u_i+1)^2-4u_i r}}{2u_i}\right],
    \label{eq:copula_generator}
\end{align}
where $u_i\equiv \rho[2F(x_i)-1]$ and $F^{-1}$ is the inverse function of $F$. The procedure is repeated until $n$ numbers are drawn. 

For the simulation, we adopt a uniform distribution $P(x)=1$ for $0\leq x<1$, implying $F(x)=x$ and $F^{-1}(x)=x$ in Eq.~\eqref{eq:copula_generator}, to generate $100$ different time series with $n=10^7$ for various values of $\rho$. Then we derive the directed horizontal visibility graphs from them to obtain the outdegree distribution for each time series and merge them to get the aggregate outdegree distribution, denoted by $\tilde P^{\rm out}_{\rho}(k)$. 

Figure~\ref{fig:DHVG} shows the numerical results of $\tilde P^{\rm out}_{\rho}(k)$ as a function of $\rho$ for $k=1,\ldots,12$. As expected from the analysis, $\tilde P^{\rm out}_{\rho}(1)$ is constant of $\rho$, while $\tilde P^{\rm out}_{\rho}(2)$ [$\tilde P^{\rm out}_{\rho}(3)$] decreases (increases) with $\rho$. For $k=2,3$, although we have assumed that $|\rho|\ll 1$, the linear solutions in Eqs.~\eqref{eq:P2_out_sol} and~\eqref{eq:P3_out_sol} are in good agreement with the numerical results up to $|\rho|\approx 0.2$ for $k=2$ and up to $|\rho|\approx 0.4$ for $k=3$, respectively. We also fit those results with the following tenth order polynomial function:
\begin{align}
    g(\rho)=\sum_{s=0}^{10} c_s \rho^s.
    \label{eq:poly}
\end{align}
Here $c_0$ corresponds to the solutions of $P^{\rm out}_0(k)$, while we are interested in estimated values of $c_1$ to compare them with the analytical results. Precisely, we estimate $c_1\approx -0.00557(2)$ for $k=2$ and $c_1\approx 0.00537(1)$ for $k=3$. These two $c_1$ values are indeed very close to corresponding analytical values, i.e., $-1/180\approx -0.00555$ for $k=2$ and $29/5400\approx 0.00537$ for $k=3$.

We explore the numerical results further by looking at $\tilde P^{\rm out}_{\rho}(k)$ for $k\geq 4$; we plot relative differences between our results and the exact solutions for the uncorrelated case in Eq.~\eqref{eq:Pk_out_uncorrel}, defined as
\begin{align}
    \Delta_{\rho,k} \equiv \frac{\tilde P^{\rm out}_{\rho}(k)- P^{\rm out}_0(k)}{P^{\rm out}_0(k)}.
    \label{eq:Delta}
\end{align}
Here we divide the difference by $P^{\rm out}_0(k)$ to plot the results of different $k$s on the same scale in the figure, which does not affect the qualitative behaviors of $\tilde P^{\rm out}_{\rho}(k)$. We find that $\tilde P^{\rm out}_{\rho}(4)$ monotonically increases with $\rho$, while $\tilde P^{\rm out}_{\rho}(5)$ decreases until $\rho\approx 0.15$, and then increases with $\rho$. Finally, all $\tilde P^{\rm out}_{\rho}(k)$s for $k\geq 6$ show monotonically decreasing behaviors with $\rho$.

\section{Horizontal visibility graph}\label{sec:HVG}

\subsection{Analysis}

In this section, we analytically derive the degree distribution of the original, undirected horizontal visibility graphs for the model time series presented in Sec.~\ref{sec:model}. The minimum degree is $k=2$. The case with $k=2$ occurs for three consecutive $x$ values, i.e., $x_{-1}$, $x_0$, and $x_1$, as long as $x_0\leq x_{-1},x_1$. Then the probability of nodes having the degree of two is written as
\begin{align}
    P^{\rm und}_{\rho}(2)&=\int_{-\infty}^{\infty} {\rm d}x_{-1} P(x_{-1})
    \int_{-\infty}^{x_{-1}}{\rm d}x_0 P(x_0|x_{-1})\notag\\
    &\times \int_{x_0}^{\infty} {\rm d}x_1 P(x_1|x_0).
    \label{eq:P2_define}
\end{align}
Using $P(x_{i+1}|x_i)$ in Eq.~\eqref{eq:Px|x}, we obtain the exact solution as
\begin{align}
    P^{\rm und}_{\rho}(2)=\frac{1}{3}-\frac{1}{30}\rho+\frac{1}{210}\rho^2
    \label{eq:P2_sol}
\end{align}
for the entire range of $\rho$ as well as for the arbitrary functional form of $P(x)$.

Next, we study the case with $k=3$, for which two situations are considered with four ordered $x$ values. The first one is depicted in Fig.~\ref{fig:scheme}(c), where $x_0\leq x_{-1}, x_2$ and $x_0>x_1$, and there can be an arbitrary number of intermediate values, say $z_1,\ldots,z_p$, between $x_1$ and $x_2$ such that $z_i\leq x_1$ for all $i=1,\ldots,p$. The second one is that $x_0\leq x_{-2}, x_1$ and $x_0>x_{-1}$, while there can be an arbitrary number of intermediate values, say $z_1,\ldots,z_p$, between $x_{-2}$ and $x_{-1}$ such that $z_i\leq x_{-1}$ for all $i=1,\ldots,p$. Thanks to the time-reversal symmetry of the FGM copula, these two cases lead to the same result. Thus we write
\begin{align}
    P^{\rm und}_{\rho}(3)&=2
    \int_{-\infty}^{\infty} {\rm d}x_{-1} P(x_{-1})   
    \int_{-\infty}^{x_{-1}}{\rm d}x_0 P(x_0|x_{-1})\notag\\
    &\times \int_{-\infty}^{x_0} {\rm d}x_1 P(x_1|x_0)
    \sum_{p=0}^{\infty}h_p(x_0,x_1)
    \label{eq:P3_define}
\end{align}
where $h_0$ and $h_p$ for $p\geq 1$ are identical to Eqs.~\eqref{eq:G0_define} and~\eqref{eq:Gp_define}, enabling us to reuse the summation of $h_p$ for all possible $p$s in Eq.~\eqref{eq:Gp_sum}. We derive the analytical solution as follows:
\begin{align}
    P^{\rm und}_{\rho}(3)=\frac{2}{9}+\frac{11}{600}\rho +\mathcal{O}(\rho^2).
    \label{eq:P3_sol}
\end{align}

For the case with $k=4$, three situations are considered with five ordered $x$ values. The first one shown in the upper panel of Fig.~\ref{fig:scheme}(d) is such that $x_0\leq x_{-1},x_3$, $x_0>x_1$, and $x_1<x_2<x_0$, while there can be an arbitrary number of intermediate values, say $z_1,\ldots,z_p$, between $x_1$ and $x_2$ such that $z_i\leq x_1$ for all $i=1,\ldots,p$, as well as an arbitrary number of intermediate values, say $z'_1,\ldots,z'_q$, between $x_2$ and $x_3$ such that $z'_j\leq x_2$ for all $j=1,\ldots,q$. The second one is in fact the time-reversed sequence of values of the first situation, leading to the same result as in the first situation. The third one is depicted in the lower panel of Fig.~\ref{fig:scheme}(d), where $x_0\leq x_{-2},x_2$ and $x_0>x_{-1},x_1$, and there can be an arbitrary number of intermediate values, say $z_1,\ldots,z_p$, between $x_1$ and $x_2$ such that $z_i\leq x_1$ for all $i=1,\ldots,p$, as well as an arbitrary number of intermediate values, say $z'_1,\ldots,z'_q$, between $x_{-2}$ and $x_{-1}$ such that $z'_j\leq x_{-1}$ for all $j=1,\ldots,q$. Then one can write
\begin{align}
    P^{\rm und}_{\rho}(4)=2R_1(\rho)+R_3(\rho),
\end{align}
where $R_1(\rho)$ and $R_3(\rho)$ denote contributions from the first and the third situations as mentioned. We first write $R_1(\rho)$ as follows:
\begin{align}
    R_1(\rho)&\equiv \int_{-\infty}^{\infty} {\rm d}x_{-1} P(x_{-1}) \int_{-\infty}^{x_{-1}}{\rm d}x_0 P(x_0|x_{-1})\notag\\
    &\times \int_{-\infty}^{x_0} {\rm d}x_1 P(x_1|x_0) \sum_{p,q=0}^{\infty} h_{pq}(x_0,x_1),
    \label{eq:R1_define}
\end{align}
where $h_{pq}(x_0,x_1)$s are identical to those in the case with $k=3$ for the directed horizontal visibility graph; using Eq.~\eqref{eq:Gpq_sum} one gets
\begin{align}
    R_1(\rho)=\frac{1}{27}+\frac{2881}{216000}\rho +\mathcal{O}(\rho^2).
\end{align}
As for $R_3(\rho)$, we again utilize the time-reversal symmetry of the FGM copula to take the integration of $x_0$, rather than the first value among $x$s, as the last integration, enabling us to write
\begin{align}
    R_3(\rho) &\equiv \int_{-\infty}^{\infty} {\rm d}x_0 P(x_0)  \sum_{p,q=0}^{\infty} h'_{pq}(x_0).
\end{align}
\begin{widetext}
Here we have for $p=q=0$
\begin{align}
    h'_{00}(x_0) &\equiv 
    \int_{-\infty}^{x_0} {\rm d}x_1 P(x_1|x_0)
    \int_{x_0}^{\infty} {\rm d}x_2 P(x_2|x_1) 
    \int_{-\infty}^{x_0} {\rm d}x_{-1} P(x_{-1}|x_0)
    \int_{x_0}^{\infty} {\rm d}x_{-2} P(x_{-2}|x_{-1}),
\end{align}
for $p\geq 1$ and $q=0$
\begin{align}
    h'_{p0}(x_0)&\equiv 
    \int_{-\infty}^{x_0} {\rm d}x_1 P(x_1|x_0)
    \int_{-\infty}^{x_1} {\rm d}z_1 P(z_1|x_1)
    \left[\prod_{i=2}^p \int_{-\infty}^{x_1} {\rm d}z_i P(z_i|z_{i-1})\right] \int_{x_0}^{\infty} {\rm d}x_2 P(x_2|z_p) \notag\\
    &\times
    \int_{-\infty}^{x_0} {\rm d}x_{-1} P(x_{-1}|x_0)
    \int_{x_0}^{\infty} {\rm d}x_{-2} P(x_{-2}|x_{-1}),
\end{align}
for $p=0$ and $q\geq 1$
\begin{align}
    h'_{0q}(x_0) &\equiv 
    \int_{-\infty}^{x_0} {\rm d}x_1 P(x_1|x_0)
    \int_{x_0}^{\infty} {\rm d}x_2 P(x_2|x_1) 
    \notag\\
    &\times
    \int_{-\infty}^{x_0} {\rm d}x_{-1} P(x_{-1}|x_0)
    \int_{-\infty}^{x_{-1}} {\rm d}z'_1 P(z'_1|x_{-1})
    \left[\prod_{j=2}^q \int_{-\infty}^{x_{-1}} {\rm d}z'_j P(z'_j|z'_{j-1})\right] \int_{x_0}^{\infty} {\rm d}x_{-2} P(x_{-2}|z'_q),
\end{align}
and for $p,q\geq 1$
\begin{align}
    h'_{pq}(x_0)&\equiv 
    \int_{-\infty}^{x_0} {\rm d}x_1 P(x_1|x_0)
    \int_{-\infty}^{x_1} {\rm d}z_1 P(z_1|x_1)
    \left[\prod_{i=2}^p \int_{-\infty}^{x_1} {\rm d}z_i P(z_i|z_{i-1})\right] \int_{x_0}^{\infty} {\rm d}x_2 P(x_2|z_p) \notag\\
    &\times
    \int_{-\infty}^{x_0} {\rm d}x_{-1} P(x_{-1}|x_0)
    \int_{-\infty}^{x_{-1}} {\rm d}z'_1 P(z'_1|x_{-1})
    \left[\prod_{j=2}^q \int_{-\infty}^{x_{-1}} {\rm d}z'_j P(z'_j|z'_{j-1})\right] \int_{x_0}^{\infty} {\rm d}x_{-2} P(x_{-2}|z'_q).
\end{align}
\end{widetext}
We observe that
\begin{align}
    \sum_{p,q=0}^{\infty} h'_{pq}(x_0)=\left[\int_{-\infty}^{x_0} {\rm d}x_1 P(x_1|x_0)\sum_{p=0}^\infty h_p(x_0,x_1)\right]^2,
\end{align}
where $h_p$s in the brackets are the same as Eqs.~\eqref{eq:G0_define} and~\eqref{eq:Gp_define}. Thus using Eq.~\eqref{eq:Gp_sum} we obtain
\begin{align}
    R_3(\rho)=\frac{2}{27}-\frac{131}{10800}\rho +\mathcal{O}(\rho^2),
\end{align}
finally leading to
\begin{align}
    P^{\rm und}_{\rho}(4)=\frac{4}{27}+\frac{1571}{108000}\rho +\mathcal{O}(\rho^2)
    \label{eq:P4_sol}
\end{align}
for any $P(x)$. In all cases with $k=2,3,4$, the analytical solutions with $\rho=0$ reduce to those for the uncorrelated case in Eq.~\eqref{eq:Pk_und_uncorrel}. Note that the analytical solutions of $P^{\rm und}_{\rho}(k)$ for $k\geq 5$ can be calculated in a similar way.

\begin{figure*}[!t]
\centering
\includegraphics[width=0.85\textwidth]{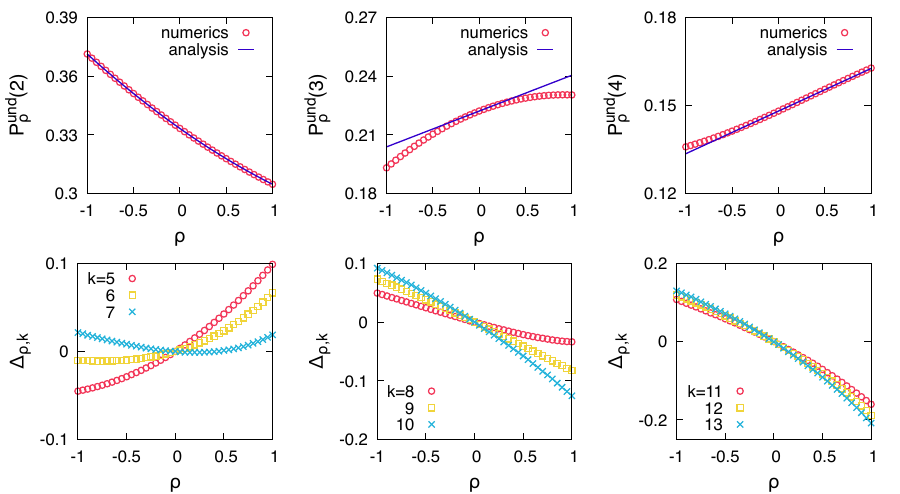}
\caption{Numerical results (symbols) of the degree distribution, i.e., $\tilde P^{\rm und}_{\rho}(k)$ for $k=2,\ldots,13$, of the horizontal visibility graphs derived from the same time series analyzed in Fig.~\ref{fig:DHVG}. In upper panels solid lines respectively show the analytical solutions of $P^{\rm und}_{\rho}(2)$ in Eq.~\eqref{eq:P2_sol}, 
$P^{\rm und}_{\rho}(3)$ in Eq.~\eqref{eq:P3_sol}, and $P^{\rm und}_{\rho}(4)$ in Eq.~\eqref{eq:P4_sol} as functions of $\rho$. In lower panels $\Delta_{\rho,k}$ in Eq.~\eqref{eq:Delta_HVG} is plotted for $k=5,\dots,13$.
}
\label{fig:HVG}
\end{figure*}

\subsection{Numerical simulation}\label{subsec:HVG_numerics}

We use the same correlated time series generated in Subsec.~\ref{subsec:DHVG_numerics} to derive the horizontal visibility graphs from them. We measure the degree distributions, denoted by $\tilde P^{\rm und}_\rho(k)$, for $k=2,\ldots,13$, as shown in Fig.~\ref{fig:HVG}. As expected, $\tilde P^{\rm und}_\rho(2)$ is in good agreement with the analytical result of $P^{\rm und}_\rho(2)$ in Eq.~\eqref{eq:P2_sol} for the entire range of $\rho$. We also find that both $\tilde P^{\rm und}_{\rho}(3)$ and $\tilde P^{\rm und}_{\rho}(4)$ increase with $\rho$ as in Eqs.~\eqref{eq:P3_sol} and~\eqref{eq:P4_sol}, where the analytical results fit well with the numerical ones up to $|\rho|\approx 0.25$ for $k=3$ and for $\rho>-0.3$ for $k=4$, despite of the assumption that $|\rho|\ll 1$. We fit the numerical results with the tenth polynomial function in Eq.~\eqref{eq:poly} to estimate the coefficients of the first order term of $\rho$ as $c_1=0.01833(1)$ for $k=3$ and $c_1=0.01456(1)$ for $k=4$, which are close to their analytical counterparts, i.e., $11/600\approx 0.01833$ for $k=3$ and $1571/108000\approx 0.01455$ for $k=4$, respectively.

We explore the numerical results further by looking at $\tilde P^{\rm und}_{\rho}(k)$ for $k\geq 5$; we plot relative differences between our results and the exact solutions for the uncorrelated case in Eq.~\eqref{eq:Pk_und_uncorrel}, defined as
\begin{align}
    \Delta_{\rho,k} \equiv \frac{\tilde P^{\rm und}_{\rho}(k)- P^{\rm und}_0(k)}{P^{\rm und}_0(k)}.
    \label{eq:Delta_HVG}
\end{align}
We find that $\tilde P^{\rm und}_{\rho}(k)$s for $k=5,6$ monotonically increase with $\rho$, while $\tilde P^{\rm out}_{\rho}(7)$ decreases until $\rho\approx 0.2$, and then increases with $\rho$. Finally, all $\tilde P^{\rm out}_{\rho}(k)$s for $k\geq 8$ show monotonically decreasing behaviors with $\rho$.

\section{Conclusion}\label{sec:concl}

By adopting the Farlie-Gumbel-Morgenstern (FGM) copula method~\cite{Nelsen2006Introduction}, we have derived the analytical results of degree distributions for $k=2,3,4$ of the horizontal visibility graph (HVG) as well as indegree and outdegree distributions for $k=1,2,3$ of the directed horizontal visibility graph (DHVG), both of which are obtained from the correlated time series. All results are obtained up to the first order of the correlation parameter $\rho$ defined in the FGM copula. These analytical results are found to be well supported by numerical simulation results. We analytically and numerically find that $\rho$ has different roles on the degree distributions of the derived HVG and DHVG; the probability of nodes having the degree $k$ can be either a constant, increasing, decreasing, or non-monotonic function of $\rho$, depending on $k$. Our analytical results shed light on exactly how the correlation between data points in the time series can affect the network structure of the HVG and DHVG.

In this work, we have considered only the correlation between two consecutive data points in the time series, while the real-world time series can show longer-range correlations beyond the pairwise correlation. To tackle such longer-range correlations, one can adopt the vine copula method~\cite{Takeuchi2020Constructing, Jo2022Copulabased} that enables to write down the joint probability distribution for more than two consecutive data points in a tractable form.

\begin{acknowledgments}
J.-M.L. and H.-H.J. acknowledge financial support by the National Research Foundation of Korea (NRF) grant funded by the Korea government (MSIT) (No. 2022R1A2C1007358).
\end{acknowledgments}



%


\end{document}